\documentclass[12pt]{article} 
\usepackage{amssymb,latexsym,amsmath}
\newtheorem{Def}{Definition}[section]

\begin{document}
\begin{center}
{\LARGE {\bf Recursive Local Fractional Derivative}} 
\vskip 1cm
{\Large {\bf Kiran M. Kolwankar}}
\vskip 1cm
{\large {\bf Department of Physics, \\
Ramniranjan Jhunjhunwala College, \\
Ghtakopar(W), Mumbai 400086\, India}}\\
\medskip 
Kiran.Kolwankar@gmail.com\\
\end{center}

\begin{abstract}
The definition of the local fractional derivative has been generalised to the
orders beyond the critical order. This makes it possible to retain more terms
in the local fractional Taylor expansion leading to better approximation.
This also extends the validity of the product rule.
\end{abstract}
\vskip 1cm
\noindent

\section{Introduction}
Though the origin of the fractional calculus~\cite{2OS,SAM,POD} is more than 300 hundred years old, the issues
related to various definitions of the fractional derivatives and integrals and their interrelation 
were resolved only towards the end of 19th century. Also, the modifications of the definitions
applicable to some specific class of problems are being introduced to date. One reason for the
diversity of definitions is the dependence of the derivative of the fractional order on the
lower limit making it a nonlocal operator as in the Riemann-Liouville definition. Taking this
limit to infinity gave rise to the Weyl's definition. In 1996, we~\cite{2KG1} introduced the
opposite limit leading to a concept of local fractional derivative (LFD). Such a derivative
turned out to be useful to characterise the local scaling behavior and the maximum order of
existence of this derivative was related to the local H\"older exponent.
The definition was further extended and applied in~\cite{2KG2,2KG3,2KG4,2KG5,2KG6}.
Several authors~\cite{LDE,CC,CCK,CCC1,CYZ,Wu,Yan,AC,BDG,CCC,KLV} have tried to take further and apply this definition.
Here we introduce further modification of this derivative in order to take care of some minor
perceived drawbacks. In the next section we give the background of definitions and then in
the later sections we introduce the generalisation and discuss its ramifications.

\section{Local fractional derivative}

The purpose of this section is to state the first definition of the LFD and study some of its
consequences using simple examples.
We begin with the Riemann-Lioville definition. In all our definitions we restrict ourselves
to the right sided derivatives and those from the left side can similarly be defined.

\begin{Def} The Riemann-Liouville fractional derivative of a function $f$ of order $q$ ($0<q<1$) is defined
as:
\begin{eqnarray}
D^q_xf(x') &=& \begin{array}{ll}
D^q_{x+}f(x'), & x'>x. 
 \end{array} \nonumber \\
&=& \frac{1}{\Gamma(1-q)}
\begin{array}{ll}
\frac{d}{dx'}\int_x^{x'}f(t)(x'-t)^{-q} dt, & x'>x. 
 \end{array}
\end{eqnarray}

\end{Def}

In~\cite{2KG1}, a concept of \emph{local fractional derivative} (LFD) was introduced in order to study the
local scaling behaviour of a function. Its definition is as follows:
\begin{Def} The local fractional derivative of order $q$ ($0<q<1$) of a function $f\in C^0: \mathbb{R} \rightarrow \mathbb{R}$
is defined as \[ {\cal{D}}^q f(x) = \lim_{x'\rightarrow x} D^q_x(f(x')-f(x)) \] if the limit exists and is finite.

\end{Def}
Here, we have subtracted the value of the function $f$ at the point $x$ which is the point of interest. 
We have also introduced a limit which makes it explicitly local. 

As an example, consider 
$f(x) = x^p$ where $0<p<1$ and $x\geq 0$. One can check~\cite{KMK2} that
the LFD of this function at $x=0$ is
\begin{eqnarray} \nonumber
{\cal{D}}^qf(0)
&=& {\left\{ \begin{array}{ll}
0 & q<p \; \mbox{or} \;\; q = p+n, \;\; n=1,2,3,...\\ \Gamma(p+1) & p=q \\ \infty & \mbox{otherwise} \\
 \end{array} \right. } \nonumber
\end{eqnarray}
and at any other point $x = x_0 > 0$ it
becomes
\begin{eqnarray} \nonumber
{\cal{D}}^qf(x_0) 
&=& {\left\{ \begin{array}{ll}
0 & q<1 \;\mbox{or} \; q=2,3,4,...\\f'(x_0) \Gamma(2) & q=1 \\ \infty & \mbox{otherwise} \\
 \end{array} \right. } \nonumber
\end{eqnarray}

One observes in this example that owing to the limiting procedure the LFD has a singular behaviour and 
the LFD is zero for orders smaller than certain \emph{critical order} and infinite for most of the orders above this order.
 This leads us to the
following definition:
\begin{Def}
The degree or the critical order of LFD of the continuous function $f$ at $x$ is defined as:
\[ q_c(x) = \sup\{q: {\cal{D}}^qf(x) \; \mbox{exists at}\; x \; \mbox{and is finite} \} \]
\end{Def}
In the example above, the critical order is $p$ at $x=0$ and 1 at other values of $x$.

In order to generalize the definition to the orders beyond one we have to subtract the Taylor expansion around
the point of interest as follows\cite{2KG2}:
\begin{Def}
The LFD of order $q$ ($N<q\leq N+1$) of a function $f\in C^0: \mathbb{R} \rightarrow \mathbb{R}$
is defined as \[ {\cal{D}}^q f(x) = \lim_{x'\rightarrow x} D^q_x\left(f(x')-\sum_{n=0}^N\frac{f^{(n)}(x)}{\Gamma(n+1)} (x'-x)^n\right) \] 
if the limit exists and is finite.
\end{Def}
With this generalized definition, the critical order in the example above gets modified.
Whereas, it has the same value $p$ at $x=0$, when $x>0$ it has value $\infty$ as the function
is analytic at all points $x>0$ and hence has a Taylor series expansion around any point $x>0$.

One peculiar aspect of this definition is that, for the function in the example above,
we can not take LFD of order greater than $p$ at $x=0$. It is not only the case with this
example, but a general aspect that it is not possible to take LFD beyond the critical
order. The purpose of this paper is to
modify the definition of the LFD so as to address this issue.

As shown in~\cite{2KG1,2KG2}, a generalisation of Taylor expansion can be
derived which involves LFD. For $N<q\leq N+1$ (provided ${\cal{D}}^q$ exists and is finite), it is given by
\begin{eqnarray}
f(x') = \sum_{n=0}^{N}{f^{(n)}(x)\over{\Gamma(n+1)}}(x'-x)^n
 + {{\cal{D}}^qf(x)\over \Gamma(q+1)} (x'-x)^q + R_q(x',x) \label{taylorg}
\end{eqnarray}
where 
\begin{eqnarray}
R_q(x',x) = {1\over\Gamma(q+1)}\int_0^{x'-x} {dF(x,t;q)\over{dt}}{(x'-x-t)^q}dt
\end{eqnarray}
and
\begin{eqnarray}
F(x,x'-x;q) = {D^q_x(f(x')-f(x))}.
\end{eqnarray}
This 
generalizes the geometric interpretation of derivatives in terms of `tangents'.
It can be shown~\cite{AC} that if the LFD in equation~(\ref{taylorg}) exists
then the remainder term $R_q(x',x)$ goes to zero as $x\rightarrow x'$.

An example of a function $f(x) = ax^\alpha + b x^\beta$ where $x\geq 0$ and $0<\alpha < \beta < 1$
was considered in~\cite{KMK2}. It was found there that the local fractional Taylor expansion with
 $q=\alpha$, the critical order of $f$, then the
second term in the equation~(\ref{taylorg}) is finite and equal to $ax^{\alpha}$ and the remainder
term yields $bx^{\beta}$. With the present definition of the LFD it is not possible to improve the
approximation by including this second term in the Taylor expansion. The modified definition which
we plan to introduce gets around this problem.

Various properties of the LFD were proved and used in~\cite{AC},~\cite{BDG} and~\cite{CCC}.
The property we are interested now is the rule of a product of two functions which was
proved in~\cite{AC}. We state it in little more generality than in~\cite{AC}. If $f$ and $g$ are two functions
having the same H\"older exponent, say $\alpha$, then, for $q\leq \alpha$
\begin{eqnarray}\label{eq:product}
{\cal{D}}^q \left(f(x)g(x)\right)= f(x){\cal{D}}^q g(x) + g(x){\cal{D}}^q f(x). 
\end{eqnarray}
The $q=\alpha$ case was proved in~\cite{AC}. But it can immediately be seen to be valid
for $q<\alpha$ too since the both sides are zero. However, as discussed in the later
sections, one runs into difficulty if one
wishes to generalise this to the product of two functions with different H\"older exponents.
Again, the new definition which we are about to introduce in the next section makes this
generalisation possible. It should be pointed out here that the conclusions of reference~\cite{VET}
are not applicable here since only differentiable functions were considered there.

\section{Recursive Local Fractional Derivative}

Here we generalise the definition of the LFD in order to be able to take LFD beyond the
critical order. 
\begin{Def}
The recursive local fractional derivative (rLFD)  of order $q$ ($q>0$) of a function $f\in C^0: \mathbb{R} \rightarrow \mathbb{R}$
is defined as \[ {\cal{D}}_r^q f(x) = \lim_{x'\rightarrow x} D^q_x\left(f(x')-f(x)-\sum_{\{\alpha\}}\frac{{\cal{D}}_r^\alpha f(x)}{\Gamma(\alpha+1)} (x'-x)^\alpha \right) \] 
if the limit exists and is finite, where the sum on the RHS is over all those $\alpha$s ($<q$) such that ${\cal{D}}_r^\alpha f$ is nonzero.

\end{Def}
This is a recursive definition in the sense that the rLFD also appears on the RHS but for orders lower than $q$. That is, to find
out the rLFD at order $q$, rLFDs of all the orders less than $q$ should be known. This leads to the concept of an order set
$\{ \alpha\}$:
\begin{Def}
The order set $\{ \alpha\}(x) = \{ \alpha_1, \alpha_2, \alpha_3, \cdots, \alpha_i, \cdots\}$ is a set of nonzero positive numbers
which are the orders at which rLFD of a given function at $x$ is nonzero.
\end{Def}
This replaces the concept of critical order of the LFD. For a real analytic function, the order set is the same as $\mathbb{N}_+$
and for a nonanalytic function, with the H\"older exponent less than one, the first entry of the order set is nothing but the critical order.
For a function $f(x) = ax^\alpha + bx^\beta$, the order set is $\{ \alpha, \beta \}$. That is, the rLFD at $x=0$ is nonzero only
for these orders and zero for all other orders.

\section{Propoerties of rLFD}

The first property we consider is that of the Taylor expansion. Using the usual
steps one can derive
\begin{eqnarray}
f(x') = f(x) + \sum_{\begin{array}{c}\{\alpha\}(x)\\\alpha < q \end{array}}
{{\cal{D}}_r^\alpha f(x)\over{\Gamma(\alpha+1)}}(x'-x)^\alpha
 + {{\cal{D}}_r^qf(x)\over \Gamma(q+1)} (x'-x)^q + R_q(x',x) \label{taylorr}
\end{eqnarray}
which is the generalisation of the Taylor series for rLFD. As it is evident
from this, one can now make the higher order approximations too. For example,
if $f(x) = ax^\alpha + bx^\beta + \sum_{i=1}^\infty a_i x^i$ is the function
of interest where $0<\alpha<\beta<1$. Its critical order at $x=0$ is $\alpha$ 
and we can not take LFD of order $>\alpha$ at $x=0$. The order set at $x=0$ is
$\{\alpha\}(0)= \{\alpha,\beta,1,2,3,\cdots\}$. Now if, in~(\ref{taylorr}), we
choose $q=\beta$ then we have two terms in the Taylor approximation with powers
$\alpha$ and $\beta$. The terms corresponding to the integer powers in $f$
contribute to the remainder term in~(\ref{taylorr}). As a result, we have better
approximation of the function than it is possible just with LFD.

Now we show that rLFD enlarges the validity of the product rule~(\ref{eq:product}).
Following the same argument as in~\cite{AC}, one can show that the product rule~(\ref{eq:product})
is valid even for rLFD for sufficiently small values of $q$. We will not go into the 
arguments but demonstrate it with the help of an example. Firstly, the product rule~(\ref{eq:product}),
even though it was proved for functions with the same H\"older exponent, can be seen
to be valid for functions with different holder exponents if $q$ is smaller than 
the lower H\"older exponents of the two. As an concrete example, let $f(x)=a_1 + b_1 x^\alpha$
and $g(x) = a_2 + b_2 x^\beta$ where $\alpha < \beta$. Now, if $q<\alpha$ the product rule~(\ref{eq:product})
is valid since then both sides are zero. It is also seen to be valid if $q=\alpha$.
We can not take $q>\alpha$ since the LFD of $f$ then does not exist. This is where
the rLFD helps in generalising the rule. The generalisation of the product rule
in which rLFD replaces the LFD is now also seen to be valid for order $\alpha < q \leq \beta$.
For, $\alpha < q < \beta$, it is trivially valid since both sides are zero and
for $q=\beta$ only one term on the RHS is nonzero. In fact, the product rule will
be also be trivially valid for a small range of orders beyond $\beta$ (depending on
the sum $\alpha+\beta$ and the second entries in the index set of $f$ and $g$, if any).

\section{Conclusions}
The concept of local fractional derivative introduced earlier has found several
applications. The original definition allows one to take LFD upto certain order
called the critical order since beyond it the LFD does not exist. Here we have
shown that a simple modification of the definition makes it possible to take
the derivative beyond the critical order. The new definition, called recursive
LFD, simply subtracts the contribution due to the LFD at the critical order 
from the function and so on. This generalised definition has some interesting
implications. One is that the local fractional Taylor expansion can now be
carried out to orders beyond the critical order. This paves the way for better
approximation schemes using the local fractional Taylor expansion. Also, the
validity of the product rule gets extended over a larger range of orders and
a larger class of functions.

%\textbf{Acknowledgements}\\

\end{document}